**SOMA: A Single-Material Organic Multivibrator Adaptive Neuron for Fully Integrated PEDOT:PSS Neuromorphic Systems**

*Nikita Prudnikov*, Yeohoon Yoon, Hans Kleemann*

N. Prudnikov, Y. Yoon, H. Kleemann
Dresden Integrated Center for Applied Physics and Photonic Materials (IAPP) and Institute for Applied Physics, Technische Universität Dresden, Nöthnitzer Straße 61, 01187 Dresden, Germany
E-mail: nikita.prudnikov@tu-dresden.de

**Abstract**

Neuromorphic electronics and spiking neural networks (SNNs) offer energy-efficient data processing, essential for real-time and edge-computing applications. In particular, interfacing and processing biological signals require devices that combine electronic performance with ionic sensitivity, which are capabilities uniquely provided by organic electrochemical transistors (OECTs). However, realizing a simple, fully integrated OECT-based neuron with rich dynamics and adaptability remains challenging. Most reported implementations rely on current-driven operation, which complicates large-scale integration and neuron–neuron coupling due to the need for precise matching of operating currents and bias voltages. Here we present a voltage-driven neuron circuit based on a multivibrator oscillator architecture, entirely fabricated from poly(3,4-ethylenedioxythiophene):polystyrene sulfonate (PEDOT:PSS). The neuron exhibits tunable adaptability through an additional control input, enabling switching between burst latency and length encoding modes. We further demonstrate a hardware-implemented two-neuron unit consisting of an inhibitory and a readout neuron, where readout activity is suppressed depending on the relative timing of the inhibitory input. Finally, we demonstrate that the fabrication process is compatible with polymer dendrite growth, enabling on-chip integration of synaptic elements on the same substrate. Owing to its structural simplicity and compatibility with a single, available material, this approach offers a scalable and accessible route toward integrated OECT-based SNNs.

## 1. Introduction

The rapid expansion of artificial intelligence has revealed the fundamental limitations of conventional computing architectures, particularly their high energy consumption and limited scalability for real-time and edge-computing tasks. These challenges have intensified the demand for energy-efficient neuromorphic systems inspired by the distributed, event-driven computation of biological neural networks.[1,2] Among the most promising frameworks are spiking neural networks (SNNs), which closely emulate the computational strategies of the brain.[3] In contrast to traditional artificial neural networks, where neuron outputs are represented by continuous activation values, SNNs encode information through discrete spikes, embedding computation in both space and time. This event-driven operation enables a more efficient representation and processing of dynamic sensory information, mirroring how biological neurons encode and transmit signals.[4]

Translating these concepts into hardware requires artificial neurons that reproduce the essential dynamics of their biological counterparts. Beyond basic functions such as threshold activation and temporal or spatial summation, more complex behaviors, such as variable spiking patterns dependent on input conditions, should also be considered when designing and interpreting device-level implementations.[5] Rather than replicating full biological complexity though, the objective is to capture essential and task-specific computational features within a compact, scalable, and energy-efficient form. A variety of hardware approaches have been explored toward this goal, including complementary metal-oxide-semiconductor (CMOS) circuits,[6,7] memristive devices,[8,9] phase-change materials,[10] and ferroelectric systems,[11] each demonstrating selected aspects of spiking or adaptation of the application in highly integrated circuits.

However, one of the key objectives in this field is the development of artificial systems capable of bridging the biological domain and the external environment, such as in brain-machine interfaces or bio-hybrid intelligent and sensing platforms. Achieving this goal requires a unified, low-power, and bio-interfaced platform whose operational characteristics are compatible with biological systems. Nevertheless, despite significant progress in this direction, realizing such an integrated solution remains challenging with the approaches described above. In this context, organic electrochemical transistors (OECTs) offer a promising route toward hardware neurons with inherent ionic–electronic coupling.[12,13] OECTs operate through ionic modulation of the organic semiconductor channel, providing a natural interface to biological systems, where neural activity is likewise governed by ionic fluxes.[14,15] The volumetric capacitance and ionic permeability of these organic mixed ionic-electronic conductors naturally give rise to time-dependent behaviors such as integration, leakage, and adaptive response.[16] Their potential biocompatibility, mechanical flexibility, and low-voltage operation enable seamless integration with soft and aqueous environments,[17,18] opening possibilities for direct interaction with living tissue and dynamic modulation by biologically relevant species.[19–21] Collectively, these properties position OECTs as a compelling platform for neuromorphic devices capable of dynamic, event-driven computation and hybrid interaction with biological systems.

In recent years, a number of studies have demonstrated various circuit architectures for OECT-based neurons.[22–25] The reported designs, however, largely relied on pre-existing circuit concepts in which the neuron response is determined solely by the input current amplitude, without incorporating additional neuron modulation mechanisms characteristic of biological systems.[26] One recent study demonstrated neuron modulation through variation of electrolyte ionic composition, offering a viable route to differentiate neurons within a network; however this approach permanently assigns neurons to fixed roles, complicating in-operando modulation.[27] Moreover, most previously reported designs rely either on two different polymers for the transistors or on ambipolar materials, with both approaches having notable limitations. Using two different polymers complicates fabrication, as the two materials must be deposited in successive steps, increasing processing time and lowering yield due to possible damage to the first-applied layer. A further drawback lies in the lack of standard materials: the polymers used across different studies vary considerably, highlighting the absence of unification in the field (Table S1).[22,23,27] By contrast, poly(3,4-ethylenedioxythiophene) polystyrene sulfonate (PEDOT:PSS) — the most widely used p-type material for OECTs — is commercially available, already extensively investigated, and proving acceptable operational stability.[28,29] Developing neural circuits based solely on PEDOT:PSS could therefore accelerate progress in organic neuromorphic electronics, as it would lower the barrier for adoption and broaden the community of potential contributors.[30–34] The main challenge, however, is that PEDOT:PSS is inherently doped, resulting in normally-on transistor behavior, which makes direct translation of earlier circuit designs impractical and requires a new architectural approach.

The design of a hardware neuron circuit can be inspired either by biology (e.g., leaky integrate-and-fire models, Hodgkin-Huxley) or classical analog and digital oscillator circuits. Among the latter, the multivibrator (MV) is one of the classical analog oscillator circuits that could serve as a fundamental building block for spiking neuron circuits and it can be implemented using only one type of transistor. The MV is based on two inverters connected with one another by feedback loop consisting of resistors and capacitors. Multivibrators are typically classified into three types: astable, monostable, and bistable.[35] The astable MV oscillates continuously between its two states under a constant voltage bias, requiring no external trigger, and thus acts as a self-sustaining oscillator. The monostable MV has one stable state and switches to an unstable one only upon receiving an external trigger pulse; it then relaxes back automatically, thus responding back with a pulse. This behavior is analogous to a canonical biological neuron that generates a single action potential once the input crosses threshold. Even though the dynamics of monostable is not rich, one of the recent studies has already shown usability of digital CMOS monostable multivibrators with varied response time as neurons for a spiking neural network.[36] Finally, the bistable MV has two stable states and remains in one until an external trigger forces a transition, making it useful for memory and digital logic functions.

In this work, we present the SOMA spiking neuron circuit in an astable multivibrator architecture entirely based on PEDOT:PSS OECTs. The design integrates two functional blocks:a two-input multivibrator with two inputs, making it a hybrid between constantly oscillating astable and single-spike monostable, and a spike transformer—ensuring voltage-level compatibility for direct neuron interconnection. By combining a constant control bias with a pulse input, the circuit exhibits rich excitable dynamics, including self-sustained oscillations, stimulus-driven bursting, and tunable excitability. In contrast to previous demonstrations that relied solely on input or electrolyte composition variations to modulate neuron behavior — approaches that limit in-operando controllability – we employ the additional input to tune neuron responsiveness to the incoming stimuli. The proposed design does not rely on weak input current that complicates network construction, but instead on robust inverter-based elements driven by voltage pulses. We show that the SOMA neuron supports two temporal information encoding schemes — latency coding and burst length coding, in contrast to the commonly employed rate encoding. Furthermore, intrinsic OECT transient dynamics give rise to short-term memory, with responses that evolve depending on recent activity. To demonstrate a pathway toward fully integrated PEDOT:PSS-based SNNs, we implement a hardware-coupled inhibitory–readout neuron pair, where the inhibitory neuron is realized using the same circuit with an adjusted input voltage bias. In this configuration, the readout neuron response is suppressed to varying degrees depending on the relative timing of the inhibitory input. We further establish compatibility between the neuron fabrication process and previously demonstrated PEDOT:$PF_6$ dendrite growth, enabling on-chip integration of synaptic elements within the same material platform. Altogether, this work establishes a unified PEDOT:PSS-based framework for neuromorphic circuit design, combining functional versatility with material simplicity to advance organic spiking neural networks.

## 2. Results

### 2.1. **SOMA neuron circuit**

The SOMA neuron circuit, shown schematically in **Figure 1a**, consists of two parts. The first functional block is a multivibrator, which is implemented using two identical inverters (single transistor layout and inverter curves are shown in Figure S1) that are interconnected via a standard RC-feedback (a resistive-capacitive) loop, while each inverter itself consists of a PEDOT:PSS OECTs and a 100 kΩ resistor. For better control of the oscillations, we selected R=100 kΩ and C=50 nF, ensuring the time constant remains sufficiently above the inverter switching time (Figure S2), thus oscillation is not limited by OECT switching time of 0.5 ms. As a standard multivibrator, it generates self-sustained oscillations within a defined bias range applied to the $MV_{input}$ node (Figure S3). The eigenfrequency of multivibrator under constant voltage bias is distributed around 50 Hz and slightly increases close to the edges of the oscillation range. The relatively fast switching of the inverter indicates that higher-frequency operation with reduced RC components is feasible. In this work, larger RC values are employed to ensure stable and reliable operation.

To extend the range of operational modes, we incorporated two separate inputs into the MV circuit labeled as $V_{control}$ and $V_{spike}$, thereby creating a hybrid configuration capable of both permanent oscillation and stimulus-initiated bursting. The first input, referred to as $V_{control}$, takes a constant voltage bias and determines the overall state of the neuron circuit. The second input $V_{spike}$ is added to receive voltage pulses from the other neurons in a network. When the neuron is stimulated by short pulse trains through this input, the MV generates a wave packet rather than self-sustained oscillations. This behavior, which will be discussed in more detail below, is indicated in the bottom of Figure 1a, where it is shown how the neurons react to a sequence of incoming pulses.

When both inputs are active, $V_{spike}$ receives stimuli from other neurons modulated by a preceding weight function element, which has previously been demonstrated using PEDOT:PSS OECTs,[16] thereby maintaining conceptual and technological consistency within an all-OECT neuromorphic architecture. To emulate the effect of the weight function when working with a single neuron, we vary the amplitude of pulses applied to this input. In a biological analogy, this amplitude represents the strength of the incoming stimulus (Figure 1b**,** bottom left). $V_{control}$, on the other hand, modulates the neuron responsiveness and sensitivity to incoming stimuli, thus being an internal neuron parameter. It can be set independently for each neuron in a network, thereby enabling simple modulation and introducing an additional degree of freedom in the training process. Biologically, this can be compared to differences in the number of ionic channels within the neuron membrane, which determines both the speed and the intensity of its response (Figure 1b**,** bottom right). The input resistors $R_{control}$ and $R_{spike}$ are both set to 10 kΩ, which is significantly lower than the resistors in the MV core to ensure efficient coupling of the applied input voltages to the MV input node.

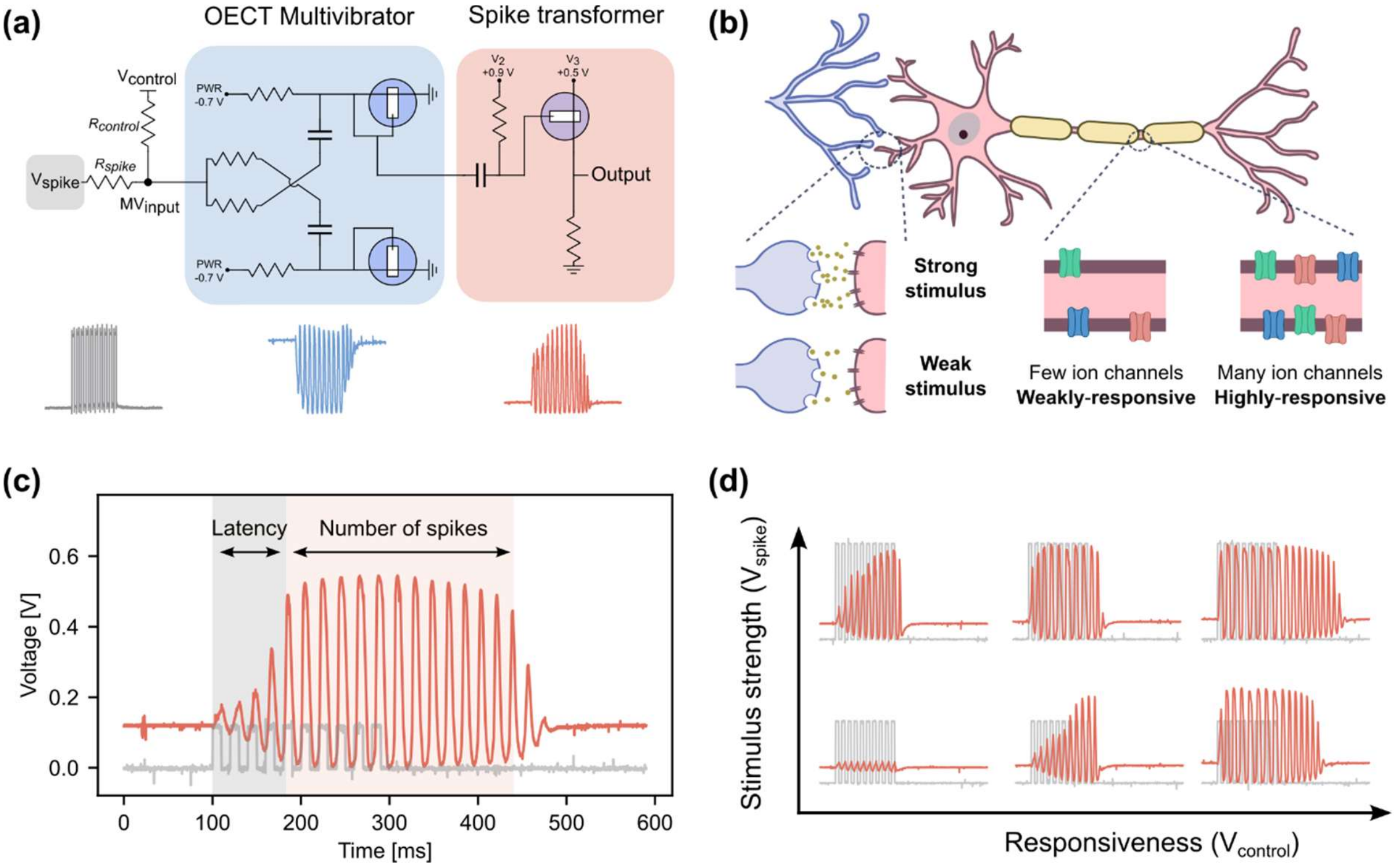

**Figure 1**. (a) Circuit diagram of the SOMA neuron based on a PEDOT:PSS OECT multivibrator core. The bottom row of the diagram includes the input signal (gray trace) and circuit response from multivibrator part (blue trace) and spike transformer (orange trace). Unmarked resistors and capacitors are 100 kΩ and 50 nF, respectively. (b) Schematic representation of a biological neuron, illustrating the analogy between stimulus strength and responsiveness modulation by ionic channels and the two inputs of the SOMA neuron. (c) Representative response of the SOMA neuron (orange trace) to 10 input pulses (gray trace), highlighting 2 response parameters: time to response and number of spikes. (d) Modulation of the neuron response under varying $V_{spike}$ (stimulus strength) and $V_{control}$ (responsiveness).

The second functional block of the circuit, the spike transformer, converts the MV output into positive pulses similar to those applied to the input. This transformation ensures compatibility between the neuron output and input, enabling integration of multiple artificial neurons into larger networks without signal mismatch. Moreover, it decouples the multivibrator cores of connected neurons, preventing them from oscillating synchronously and ensuring one-directional signal propagation.

For simplicity and architectural coherence, the subcircuit is implemented using the same RC components and inverter configuration as the MV core. As a result, the complete neuron circuit consists of three structurally similar and interconnected building blocks, each composed of one OECT, two resistors, and one capacitor. As a result, simulation of the circuit behavior, e.g. with Simulation Program with Integrated Circuit Emphasis (SPICE), is very straightforward. To perform the simulation, we employ a standard PMOS element with a threshold voltage of 0.4 V, together with resistors and capacitors. This model reproduces the key circuit behavior and shows that adjusting the ratio of the multivibrator input resistors $R_{control}$ and $R_{spike}$ enables tuning of optimal control and spike amplitude range (Figure S4). By adjusting the inverter connections and applied bias voltages in the spike transformer subcircuit, the spike polarity and offset can be tuned according to the desired neuronal response (Figure S5). Moreover, this uniform architecture enables a direct estimation of power consumption based on the identical inverter resistors. Because the primary power consumption arises from current flowing through the three identical inverter resistors, the total power consumption of the neuron circuit can be estimated. For a multivibrator and spike transformer powered at −0.8 V and +0.5 V, respectively, the power consumption is 15 μW.

Figure 1c presents an example of the circuit response to an incoming pulse train. Two key features of the response can be controlled through the interaction between the two inputs and thus exploited for information encoding (Figure 1d). The first is the response latency, defined as the time delay before the circuit initiates spiking activity after the onset of stimulation. The second is the response length, expressed as the total number of spikes in the output train, which reflects the intensity and persistence of the neuron reaction (more details in the Information Encoding section). Based on the response length, four general classes of behavior can be identified (Figure S6). For weak input pulses and low control voltage, the circuit remains silent, exhibiting no response, as the combination of two voltages was not sufficient for the MV to start oscillating, due to certain amount of leakage through the inverter transistors. For stronger stimuli, it demonstrates a continuous shift towards suppression (fewer spikes than in the input), propagation (a similar number of spikes), and amplification (more spikes than in the input). Together, these controllable response modes illustrate the circuit's potential for flexible information processing, including signal filtering, gain control, and temporal coding within artificial spiking neural networks.

An extensive analysis of neuron behavior under various combinations of inputs is presented in **Figure 2a-b**, which shows the neuron response latency and response to trains of 10 voltage pulses of 50 Hz and respective $V_{control}$ and $V_{spike}$ amplitudes. Furthermore, based on these plots we can determine the threshold condition for spiking and the transition voltages for suppression, propagation and amplification of the input pulses. The numbers inside the heatmap cells provide exact values of the respective response parameters (either spike latency in ms or number of spikes in a burst), while white areas without numbers cover the areas, where the stimulation is too weak to excite the neuron. For highest control voltages (0.88 and 0.90 V), the SOMA-neuron exhibits a distinct behavior. Within a certain voltage range, it oscillates continuously, independent of applied input spikes (Figure S7). Neither of the input parameters alone determines response characteristics, rather, the response emerges from their combined effect. Nevertheless, for higher spike amplitudes, $V_{control}$ predominantly determines the number of spikes in response, e.g. higher values prolong the response for some time after input spikes ceased. The response latency, in contrast, is influenced by both input parameters.

### 2.2. **Information Encoding**

In Figure 2c, we illustrate two distinct encoding strategies available in the SOMA-neuron, both of which rely on the controllable parameters of its response. The first strategy is latency encoding, also referred to as the time-to-first-spike approach. To demonstrate the circuit ability to implement this scheme, we apply a train of 10 input pulses of varying amplitude with 50 Hz frequency and 50% duty cycle, while maintaining a constant control voltage of 0.84 V. The resulting responses are shown in Figure 2d. As evident from the traces, increasing the amplitude of the input pulse systematically reduces the delay between stimulus onset and the first output spike. In this way, the temporal position of the first spike directly reflects the stimulus strength.

The second strategy, obtained here under control voltage of 0.86 V and larger input pulses, is burst length encoding, where information is carried by the response length, i.e., the number of spikes in the output train rather than the delay of the first spike after excitation. As shown in Figure 2e, stronger input pulses elicit longer spike trains, with the number of output spikes increasing monotonically with pulse amplitude. This mode of encoding thus translates the stimulus intensity into the degree of spiking activity.

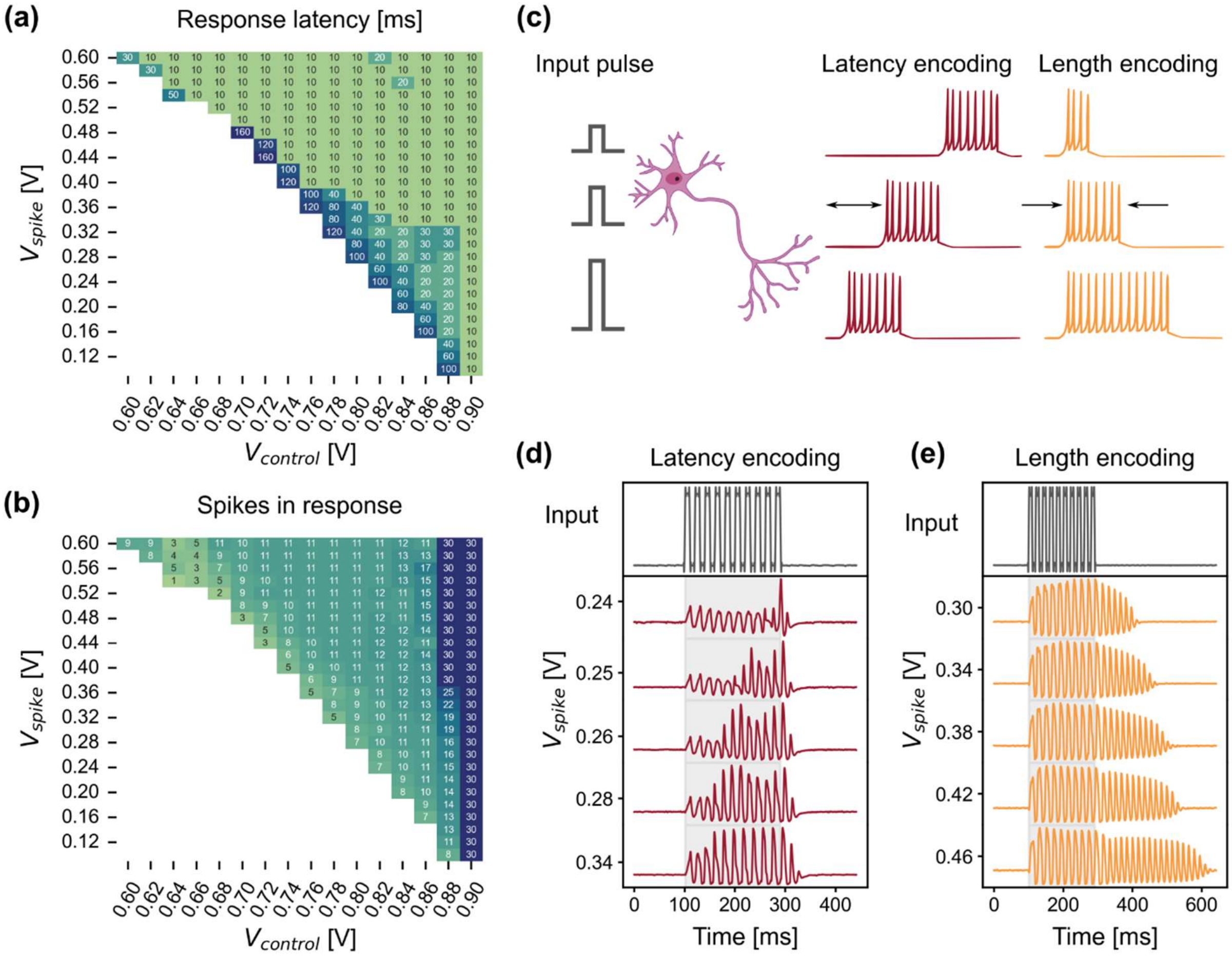

**Figure 2.** (a,b) Heatmaps of neuron response latency in ms (a) and burst length (b) over a broad range of $V_{control}$ and $V_{spike}$ values. A train of 10 pulses with 50 Hz frequency and 50% duty cycle is applied to the $V_{spike}$ input. The numbers indicate latency (in milliseconds) and the number of spikes in the response. (c) Schematic illustration of two types of information encoding in the SOMA neuron: response latency and length encoding. (d) Neuron responses to varying input spike amplitudes at $V_{control}$ = 0.84 V, showing changes in latency for different spike amplitudes. (e) Neuron responses at $V_{control}$ = 0.86 V, exhibiting variations in burst length for different spike amplitudes.

Notably, the operational voltage range can be selected to emphasize a specific encoding mechanism. For example, in the regime used for length encoding, the responses exhibit very little variation in latency, making this range effectively dedicated to spike-count–based encoding. In this case, applying ten input pulses is not necessary, as the only varying parameter is the number of output spikes generated after stimulation, which can already be modulated using a single input pulse (Figure S8). In contrast, when response latency is of primary interest, an operation at lower voltages is required. In this regime, multiple input pulses are necessary to introduce more available output response latency values. Intermediate voltage ranges can also be selected to achieve a controlled interplay between latency and spike-count modulation, thereby enabling mixed encoding strategies. This tunable behavior is reminiscent of biological neural circuits, where information processing does not rely on a single coding paradigm. Instead, structurally similar neurons across different brain regions employ distinct yet complementary encoding mechanisms that collectively support distributed information processing within a unified network.[37]

### 2.3. **Inhibitory connection and short-term memory**

Spiking neural networks comprise not only excitatory neurons that stimulate subsequent neurons to generate spikes, but also of inhibitory neurons, which suppress spiking activity in target neurons when activated. Such neurons help stabilize SNNs by balancing strong excitation and preventing network activity from diverging.[38] To illustrate how such inhibitory neurons can be implemented and operate within an SNN composed of SOMA neurons, we construct a simple two-neuron network unit (**Figure 3a**). The unit consists of a readout neuron with excitatory and inhibitory spike inputs, and an inhibitory neuron, whose output is physically connected to the inhibitory input of the readout neuron (Figure 3b). The inhibitory neuron shares the same structural design as the readout neuron; however, the voltage bias levels of its spike transformer is adjusted to produce a high positive baseline when inactive and transient downward voltage pulses upon stimulation (Figure 3b, middle, blue trace). These downward pulses, while remaining at positive potential, are sufficient to switch an additional OECT into the conducting state when the inhibitory neuron is active. This auxiliary OECT is connected between the MV input of the readout neuron and ground. When activated, the inhibitory neuron drives this OECT, enabling a voltage drop at the MV node and thereby suppressing spike generation in the readout neuron.

To illustrate the influence of the inhibitory neuron on the unit response, we applied two equal sets of five input pulses (50 Hz, 50% duty cycle) with a controlled delay $\Delta t$ between them to excitatory and inhibitory inputs. Excitatory pulses are sent directly to the excitatory input of the readout neuron, while inhibitory pulses are applied with a delay to the inhibitory neuron, which in turn generates a response transmitted to the inhibitory input, i.e. the additional OECT gate, of the readout neuron. The results showing a number of spikes of a response of the readout neuron are presented in Figure 3c. Initially, the output neuron operates in an amplification mode, producing more spikes than it receives at the excitatory input. For large *Δt* values, the inhibitory input has no effect on the output of the readout neuron, as the inhibitory spikes arrive after the output response is over (Figure 3c, “Not inhibited” panel). However, as the delay decreases and inhibitory spikes arrive within the readout neuron's active response window, its activity is partially suppressed and the neuron generates fewer spikes than in the absence of inhibition (Figure 3c, "Partly inhibited" panel). When the inhibitory input substantially overlaps with the excitatory one ($\Delta t \leq 0$), the activity of the readout neuron is fully suppressed, resulting in the absence of output spikes (Figure 3c, "Fully inhibited" panel). Thus, by gradually reducing the inhibitory delay down to zero, it is possible to shift the readout neuron behavior from amplification to propagation and ultimately to suppression mode. Additional output traces of the readout neuron for various delay values $\Delta t$ are presented in Figure S9.

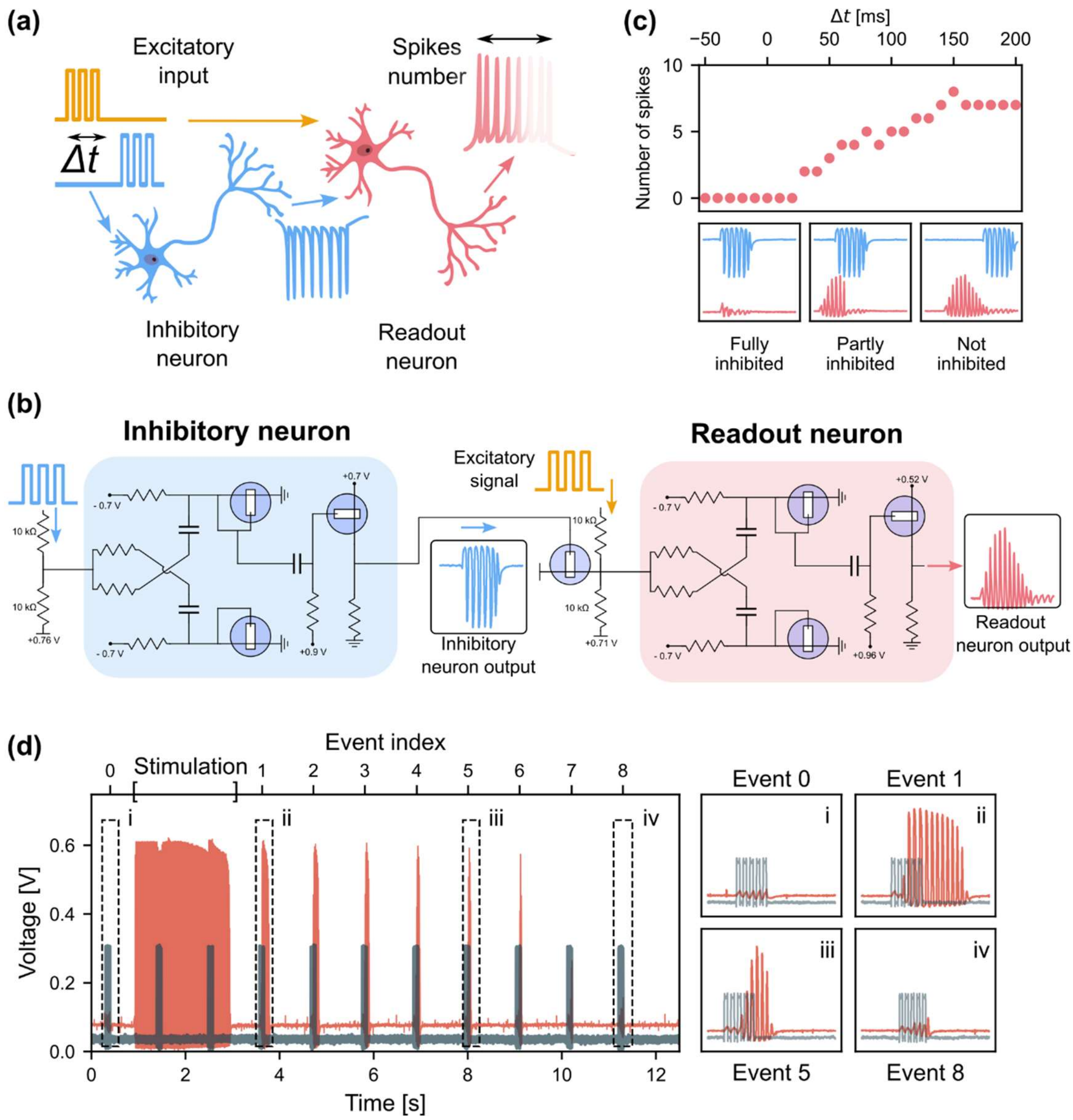

**Figure 3.** (a) Schematic illustrating the interaction between excitatory and inhibitory stimuli. The excitatory stimulus is applied to the corresponding input of the readout neuron, whereas the inhibitory stimulus drives an inhibitory neuron whose output is connected to the inhibitory input of the readout neuron. The output of the readout neuron depends on the delay (Δt) between the excitatory and inhibitory stimuli. (b) Detailed circuit diagram showing the connection between the inhibitory and readout neurons. (c) Dependence of the spike count (readout neuron biased to operate in bursting-encoding mode) in the neuronal response on the delay Δt, revealing a gradual transition from amplification to propagation and suppression modes. (d) Temporal response of the neuron (orange trace) to repetitive stimulation consisting of five consecutive 0.3 V pulses, recorded before and after conditioning with a strong stimulus, demonstrating a short-term memory effect. Bottom panels show magnified views of the highlighted regions in the upper plot.

Besides different encoding methods, the SOMA neuron also exhibits short-term memory properties, like biological neurons, which is considered to contribute to working memory alongside synaptic plasticity.[39] In Figure 3d (left panel) the evolution of neuron response for the same input sequence of five pulses (50 Hz, 50% duty cycle) in time is shown. Initially, the neuron demonstrates no response to a stimulus (Figure 3d, Event 0) as the stimulating pulses are too weak. Then the neuron is switched to constant oscillation mode for 2 s by increasing the control voltage from 0.75 to 0.95 V for 1 s, so that voltage at the MV input remains within the threshold window, regardless of whether there are incoming pulses or not (Figure S10). Then, the control voltage is lowered back to initial value of 0.75 V, under which the neuron previously demonstrated no response. However, right after the constant oscillation mode, the neuron exhibits strongly amplified response to the incoming pulse sequence (Figure 3d, Event 1). After a few cycles the response gradually decreases to zero (Figure 3d, Events 5 and 8), which is reproducible after further stimulations (Figure S11).

The effect is attributed to complex OECT transient behavior including long-term effects due to multiple time constants of the devices rather than circuit behavior.[40] To demonstrate this, we swept the OECT-inverter with a ramp signal of the same amplitude but different offset voltage. The system demonstrates gradual decrease of output voltage range and peak width in the time interval of few seconds, similar to short-term memory time of the neuron (Figure S12). As the ionic time constant is much larger than switching speed, the ionic equilibrium can be reached only after few switching cycles, thus the range of the operational voltage slowly moves towards either direction. Moreover, the described effect is not observed in SPICE modeling with MOSFET (metal–oxide–semiconductor field-effect) transistors instead of OECTs (Figure S13). Such behavior of the SOMA neuron allows better simulation of biological neurons, which demonstrate similar behavior of gradual suppression of response intensity.[41]

### 2.4. **Fully PEDOT:PSS SOMA neuron, limitations and future directions**

To demonstrate the feasibility of constructing an entire neural network using PEDOT:PSS, we first assemble a neuron circuit composed entirely of PEDOT:PSS-based components, including passive elements such as capacitors and resistors. The circuit is fabricated on a single substrate using a two-step process: spin-coating and patterning of PEDOT:PSS, followed by inkjet printing of the electrolyte (**Figure 4a**). The capacitors are formed by two PEDOT:PSS-coated interdigitated gold electrodes and covered with a printed solid-state electrolyte (Figure 4b). The same electrolyte is used for both the OECT (Figure 4c) and the capacitor, and is inkjet-printed in a single step. The resulting capacitance according to an impedance spectroscopy is approximately 34 nF (Figure S14). Resistors are implemented as patterned PEDOT:PSS lines with geometries chosen to yield approximately 100 kΩ resistance (Figure 4d). A 25×25 $mm^2$ glass substrate is used to accommodate 63 fully integrated PEDOT:PSS neurons (Figure 4e). The effective footprint, excluding contact pads, is approximately 2 $mm^2$ (Figure 4f), indicating that a substantially larger number of neurons could be integrated on the same substrate in network configurations without individual contact terminals for each unit.

As a potential solution, we identify electropolymerized conductive polymer dendritic structures as a promising approach (Figure 4g). These dendrites can be grown and operate on the same substrate as used for the neuron fabrication, thus proving the compatibility of the two technologies, supporting the feasibility of fully integrated, single-material neuromorphic systems (Figure S15**)**. In future network implementations, dendrites can be grown *in situ*, for example between simultaneously active but unconnected neurons. This enables the direct hardware implementation of the "fire together, wire together" principle, thereby mimicking synaptogenesis through activity-dependent formation of conductive pathways.[42] Moreover, dendrites composed of PEDOT modulate their conductivity after growth in response to applied voltage pulses of an amplitude an frequency compatible with the SOMA neuron (Figure S16). This behavior provides a physical basis for synaptic plasticity, as the effective coupling strength between neurons can be dynamically tuned through electrical stimulation. Together, these properties suggest that electropolymerized conductive polymer dendrites could serve not only as structural interconnects but also as a part of an adaptive synaptic element within fully integrated spiking neural networks.

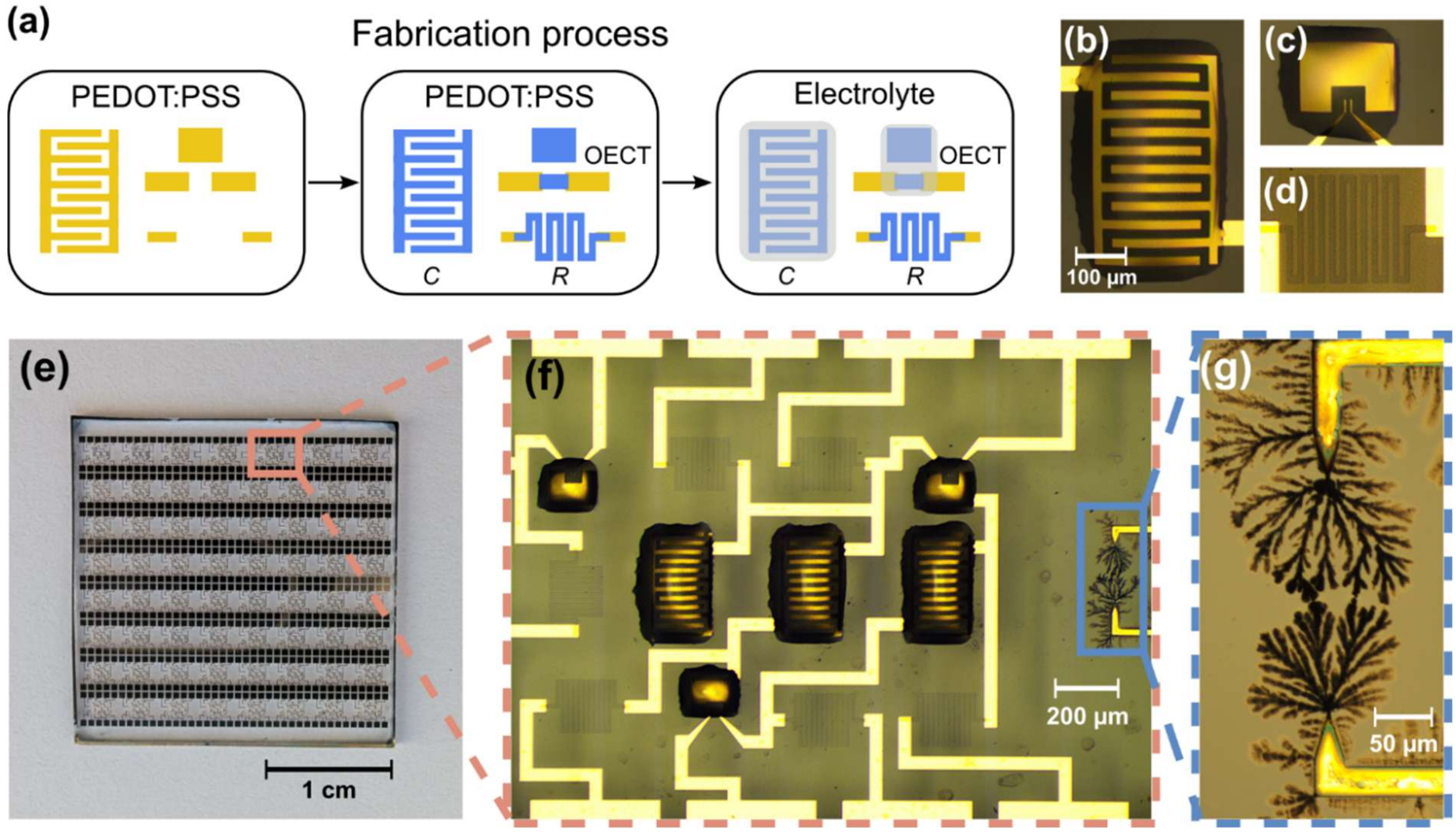

**Figure 4.** (a) Two-step fabrication process of the fully-PEDOT:PSS SOMA neuron circuit: patterning of the polymer layer followed by inkjet-printing of the electrolyte. (b–d) Optical micrographs of a PEDOT:PSS-based capacitor (b), OECT (c), and resistor (d) for the fully integrated SOMA neuron circuit, all shown with a common scale bar. (e) Glass substrate hosting 63 fully integrated neurons. (f) Optical micrograph of a fully integrated SOMA neuron circuit with an electropolymerized PEDOT:$PF_6$ dendrite (g).

### 3. Conclusion

In conclusion, we demonstrated a spiking SOMA neuron circuit based on a multivibrator architecture, implemented entirely with OECTs fabricated from the widely available material PEDOT:PSS. The design, composed of three identical blocks each containing an inverter, resistor, and capacitor, is structurally robust and requires only two fabrication steps. The self-oscillating multivibrator core, combined with the two-input configuration (a constant control bias and pulsed stimuli), enables independent and dynamic tuning of neuronal responsiveness and rich, stimulus-driven dynamics. At the same time, two separate inputs enable rapid and straightforward switching between two information encoding modes: latency and burst-length. The second stage of the circuit is a spike transformer, which decouples the oscillating core from the output and ensures voltage-level compatibility between input and output, thus making the direct connection of the neurons possible.

Beyond single-neuron operation, we demonstrated functional reconfigurability by converting the same architecture into an inhibitory neuron through adjustment of the input bias conditions. By coupling inhibitory and excitatory neurons into a small unit, we showcased a pathway toward the realization of SNNs, where inhibitory neurons play a crucial regulatory role. We showed the readout neuron to be suppressed by different degree depending on the arrival timing of inhibitory neuron activity, progressively transitioning from signal amplification to complete response suppression.

In addition to the potential biocompatibility and ease of fabrication, implementation of the circuit with OECTs enabled short-term memory of the neuron in the seconds time range. Specifically, the neuron shows gradual decrease of response down to full silence to the same small input pulses after a period of strong stimulation. This behavior arises from intrinsic OECT transient dynamics, governed by distinct ionic and electronic time constants.

Furthermore, we demonstrated a fully integrated neuron in which all active and passive components are realized exclusively using PEDOT:PSS. The compact effective footprint of approximately 2 $mm^2$ enables high-density integration, making the platform well-suited for scalable organic spiking neural networks. At the network level, we proposed an approach for realizing both excitatory and inhibitory connectivity and demonstrated that PEDOT:$PF_6$ dendritic structures are compatible with the fabrication process, paving the way for their implementation as synaptic connections between neurons. Importantly, the realization of fully integrated neuron arrays on a single glass substrate, achieved in only two fabrication steps and without heterogeneous material integration, highlights the scalability and manufacturability of the proposed platform for large-area neuromorphic systems.

This scalable and modular approach, based on a well-characterized and readily available material system, provides a practical route toward organic neuromorphic circuits, bridging fundamental research and real-world implementation. Altogether, this work establishes PEDOT:PSS-based multivibrator neurons as versatile and integrable building blocks for future organic spiking neural systems.

## 4. Methods

*OECT fabrication*: The electrodes were patterned by photolithography on glass substrates coated with 3 nm of chromium for adhesion and 50 nm of gold. Photolithography was performed by µMLA Maskless Aligner (Heidelberg Instruments, Germany) using the positive photoresist AZ1518 (MicroChemicals GmbH, Germany). Following electrode patterning, PEDOT:PSS (Heraeus, Germany) spin-coating and subsequently patterned by photolithography using the negative photoresist OSCoR 4020 (Orthogonal Inc., USA). The OECT featured a channel width, length and gate distance of 30, 5 and 20 µm, respectively. After patterning of PEDOT:PSS, five layers of solid-state electrolyte were applied to OECTs and PEDOT:PSS capacitors by the Dimatix Materials DMP-2800 inkjet printer (Fujifilm Dimatix Inc., USA), followed by UV exposure for 120s for ink solidification. The process was previously described in greater detail elsewhere.[43,44] The fabricated devices were stored and measured in a glove box under nitrogen atmosphere.

*Dendritic structures growth*: The solution for electropolymerization was prepared by dissolving 50mM of EDOT (Tokyo Chemical Industry Co., Ltd., Japan) monomer and 1 mM of $TBAPF_6$ in acetonitrile. Dendrite growth was performed by placing a droplet of the solution between two gold electrodes with a distance of 200µm, and applying 25 Hz sinusoidal voltage signal with peak-to-peak amplitude of 6 V. The voltage was maintained until the conductive dendrite bridge formed between the electrodes, which typically required 5-10 s. The growth was performed before inkjet-printing of the solid-state electrolyte.

*Electrical measurements and data analysis*: Electrical measurements were performed using a USB data acquisition card USB-1208HS-4AO (Digilent Inc., USA) controlled by custom developed Python software. Impedance measurements were carried out with a Metrohm Autolab PGSTAT302N potentiostat/galvanostat (Metrohm AG, Switzerland).

*SPICE simulations*: SPICE simulations were performed with PySpice v1.5 with Ngspice v34 as the simulation engine.

**Acknowledgements**

The authors are grateful for funding provided by the Deutsche Forschungsgemeinschaft through the project ArNeBOT (KL 2961/12-1). The authors thank Ali Solgi for valuable guidance on inkjet printing techniques, Fabian Knüpfer for developing the DAQ card library used in this study, Richard Kantelberg for insightful discussions on fibers and assistance with the use of experimental tools, Tim Köpsell for support with sample fabrication. Some graphical elements were adapted from Wikimedia Commons under the CC BY-SA 3.0 license.

**Data Availability Statement**

The data that support the findings of this study are available from the corresponding author upon reasonable request.

**Conflict of Interest**

The authors declare no conflict of interest.

# Supporting Information

**SOMA: A Single-Material Organic Multivibrator Adaptive Neuron for Fully Integrated PEDOT:PSS Neuromorphic Systems**

*Nikita Prudnikov*, Yeohoon Yoon, Hans Kleemann*

**Table S1.** Hardware OECT-based implementations of artificial spiking neuron.

| Architecture | Materials used | Encoding | Possibility of neuron stacking | Fully-integrated | Power consumption | Footprint (in case of integration) | Reference |
|---|---|---|---|---|---|---|---|
| Axon-Hillock | P(g42T-T), BBL | rate | - | yes | 15 µW | >10 cm$^2$ | [S1] |
| Negative Differential Resistance | p(g2T-TT), PEDOT:PSS | rate | - | no | 24 µW | - | [S2] |
| Hodgkin–Huxley | BBL | multiple | - | no | 60 µW | - | [S3] |
| Axon-Hillock | PBBT-Me, BBL | rate | - | - | - | - | [S4] |
| Hodgkin–Huxley | BBL, BBL/PEDOT bilayer | rate | - | no | 8.2 µW | - | [S5] |
| Negative Differential resistance | BBL | rate | - | yes | 16.43 nW | ~180 µm$^2$ | [S6] |
| Axon-Hillock | gDPP-g2T, Homo-gDPPTz | rate | - | yes | - | <37 mm$^2$ | [S7] |
| Multivibrator | PEDOT:PSS | TTFS, burst-length | yes | yes | ~ 15-20 µW | 2 mm$^2$ | this work |

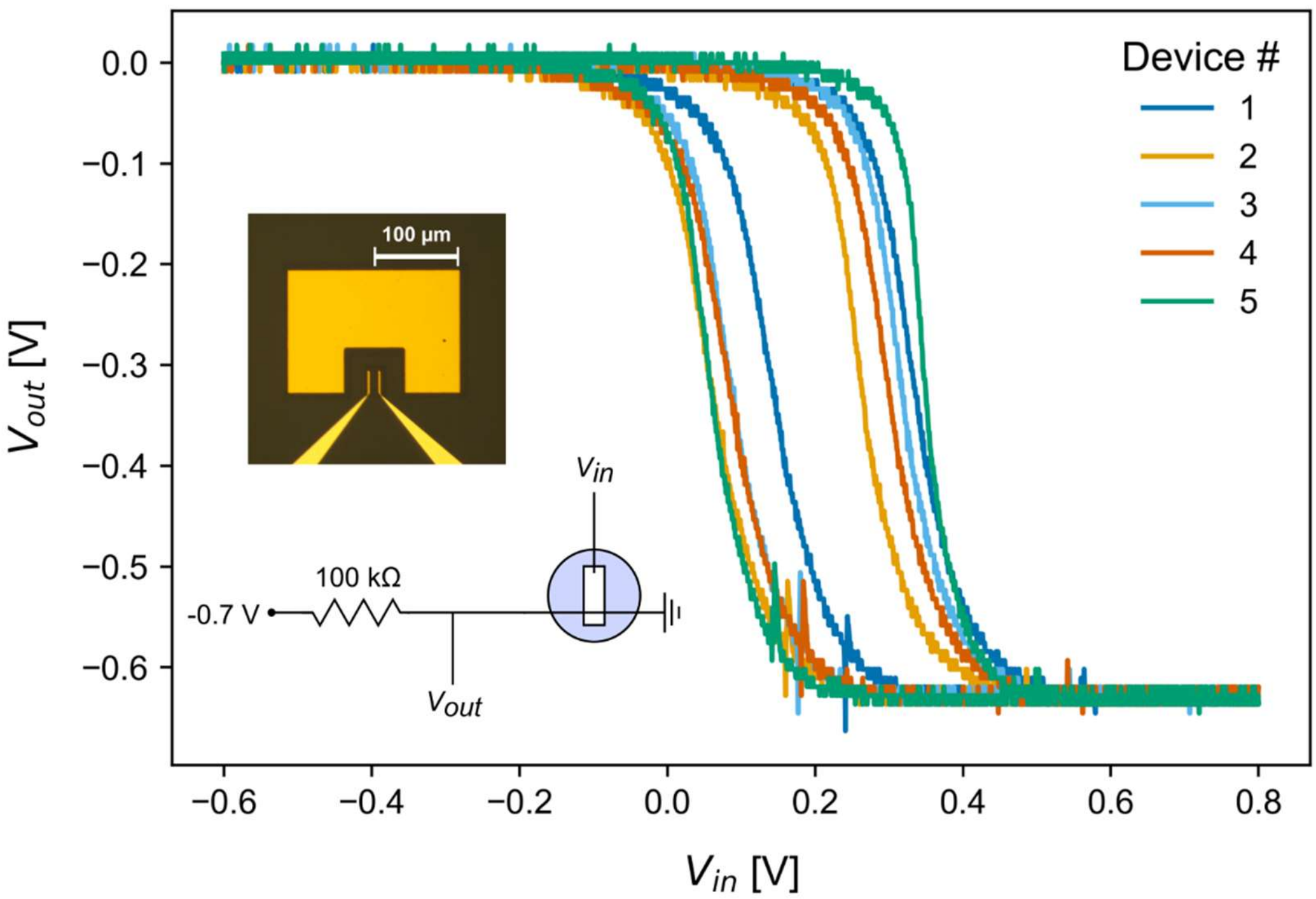

**Figure S1.** Switching curves of five OECT-based inverters, with OECTs fabricated on a single substrate. The connection diagram is shown in the inset. The inset also presents the OECT layout, with channel width and length of 30 µm and 5 µm, respectively.

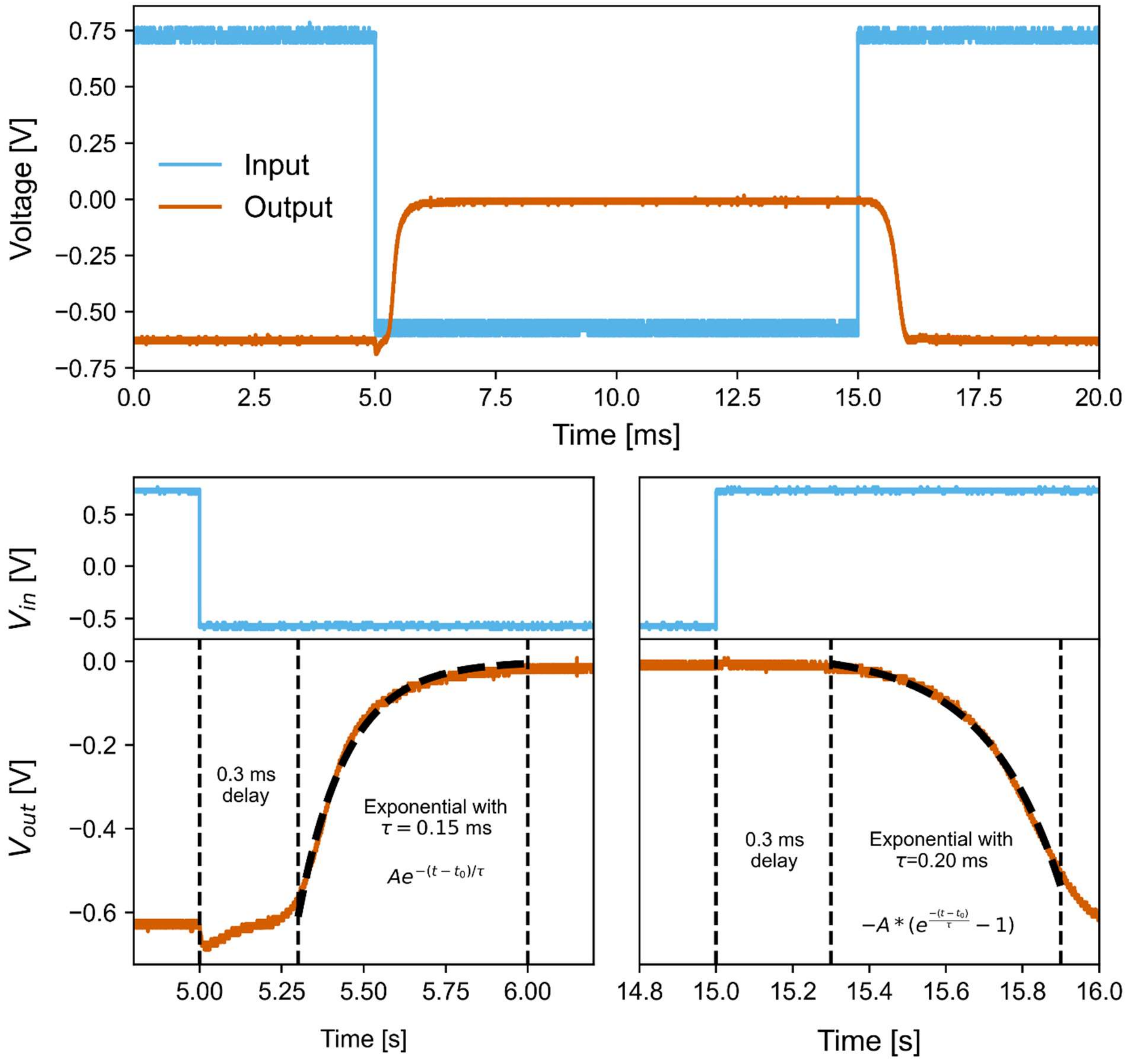

**Figure S2.** Switching behavior of the OECT inverter (orange trace) in response to 100 Hz voltage pulses (blue trace) applied to the input (top). Switching to off (bottom left) and on (bottom right) states with fitting by exponential functions (black dash line).

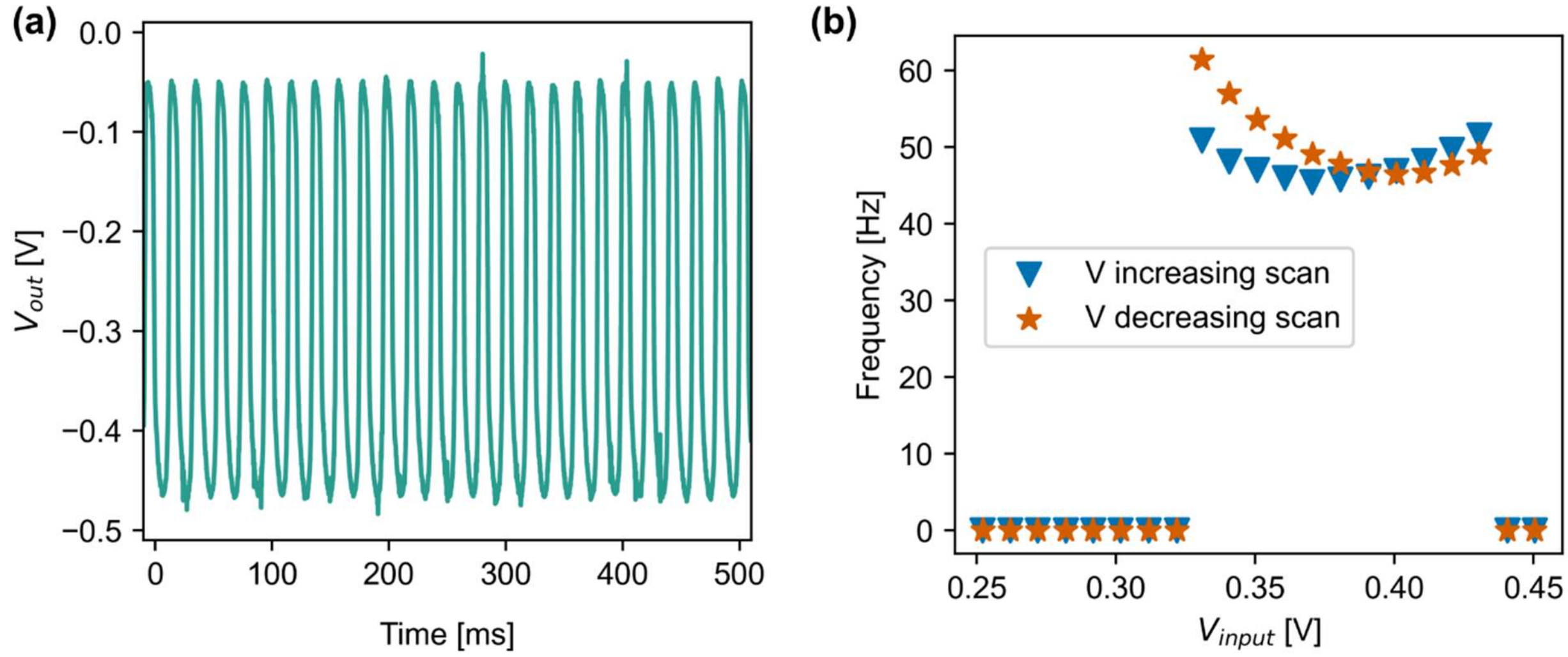

**Figure S3.** (a) Stable self-sustained oscillations of the OECT-based multivibrator under a constant voltage bias of 0.36 V. (b) Dependence of the oscillation frequency on the input voltage $V_{input}$, measured during forward and reverse voltage sweeps. Each voltage step was applied for 5 s.

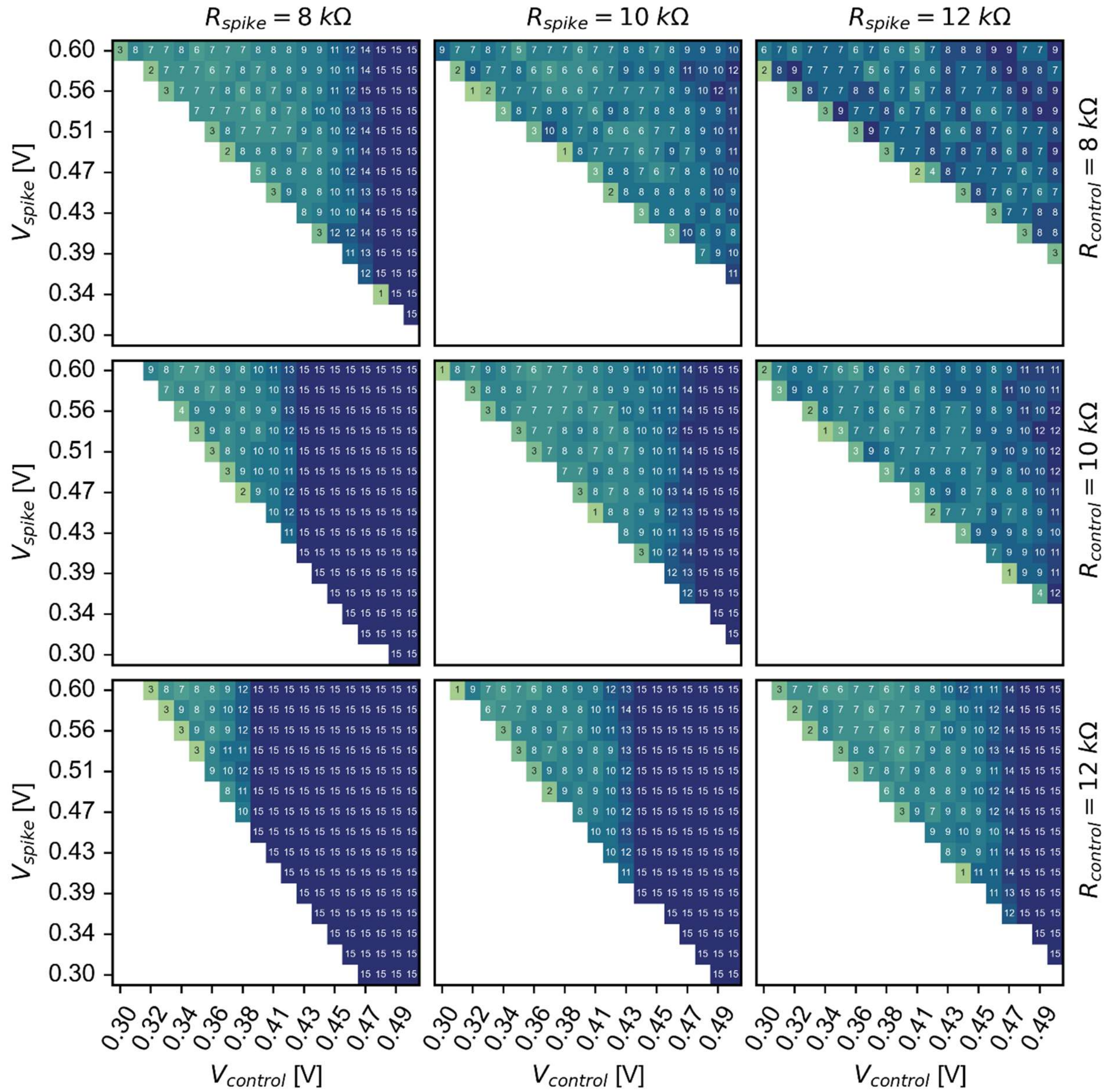

**Figure S4.** SPICE Simulated response lengths of the neuron obtained from SPICE modeling for 9 combinations of $R_{control}$ and $R_{spike}$. Stimulation was applied as a train of 10 pulses at 100 Hz with a 50% duty cycle.

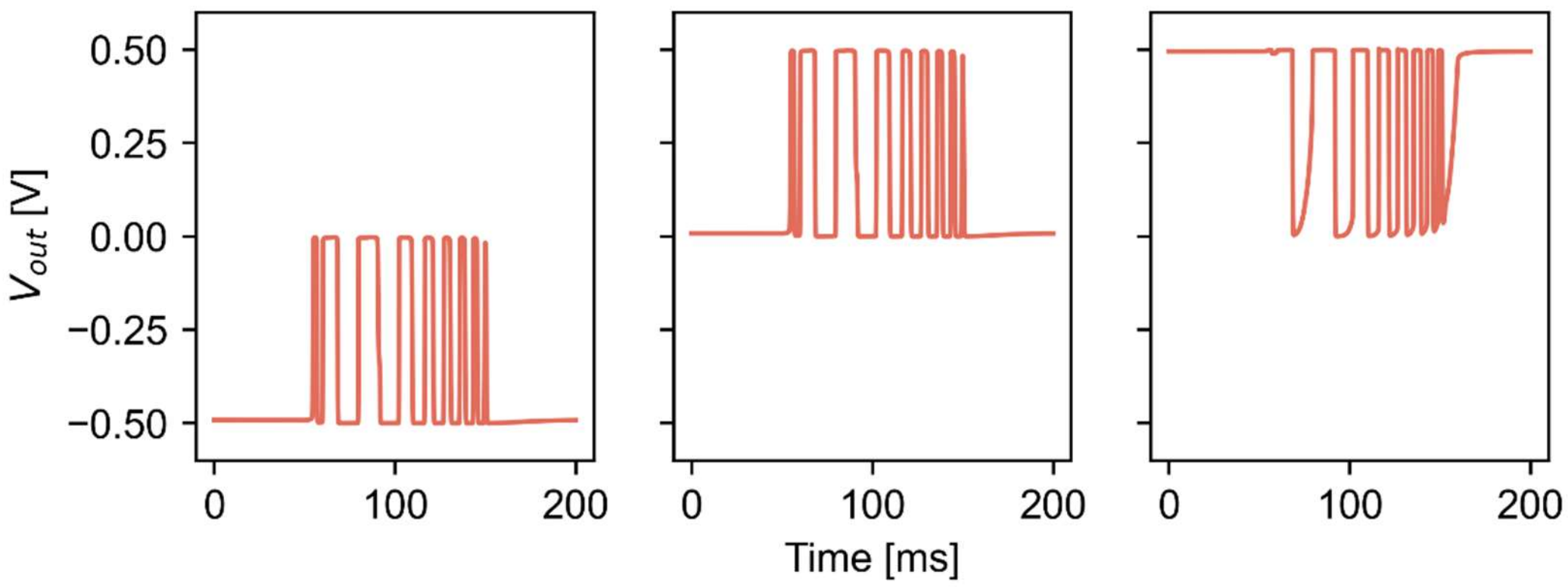

**Figure S5.** SPICE simulations of available spiking ranges for different voltages applied to Spike Transformer. Left: $V_2 = 0.5$ V, $V_3 = -0.5$ V; middle: $V_2 = 1.0$ V, $V_3 = 0.5$ V; right: $V_2 = 0.8$ V, $V_3 = 0.5$ V.

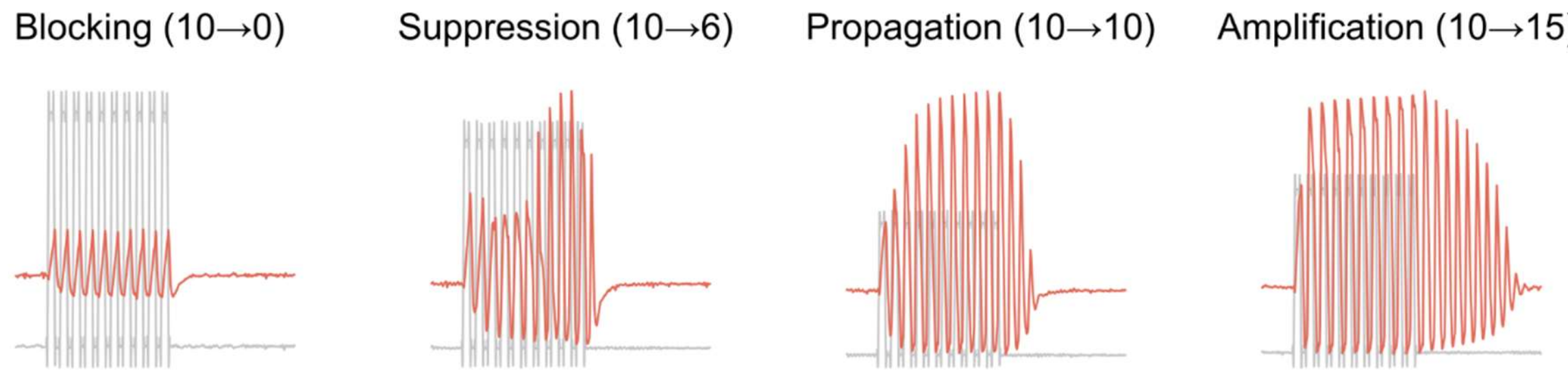

**Figure S6**. Demonstration of four types of neuron response to 10 input pulses from left to right: blocking (silent, $V_{control}$ = 0.7 V, $V_{spike}$ = 0.32 V); suppression (6 output spikes, $V_{control}$ = 0.78 V, $V_{spike}$ = 0.32 V); propagation (10 output spikes, $V_{control}$ = 0.86 V, $V_{spike}$ = 0.20 V); amplification (15 output spikes, $V_{control}$ = 0.88 V, $V_{spike}$ = 0.24 V).

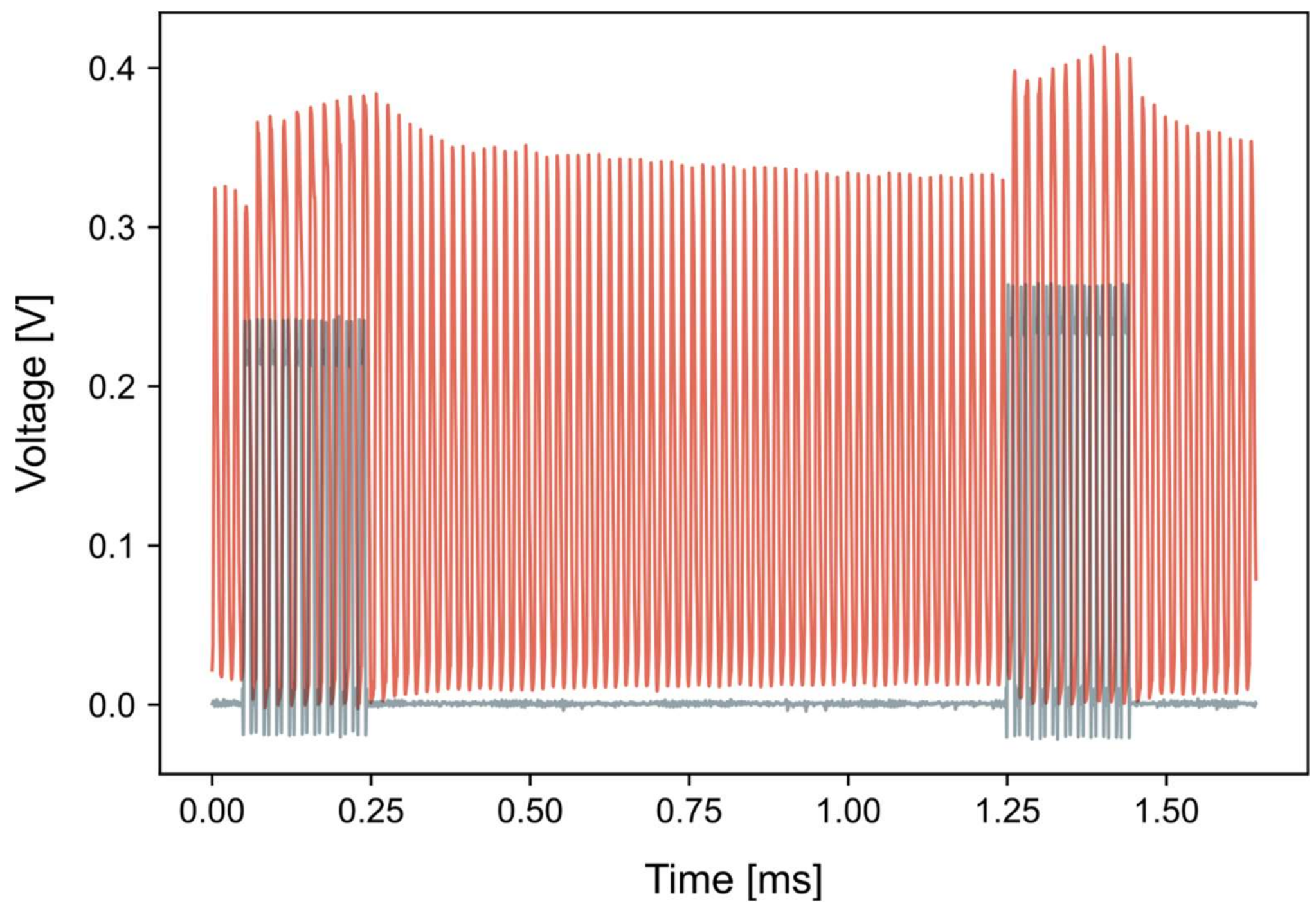

**Figure S7**. Continuous spiking of the SOMA neuron (orange trace) between sets of 10 input pulses of 0.22 V and 0.24 V (gray trace) and $V_{control}$ = 0.89 V.

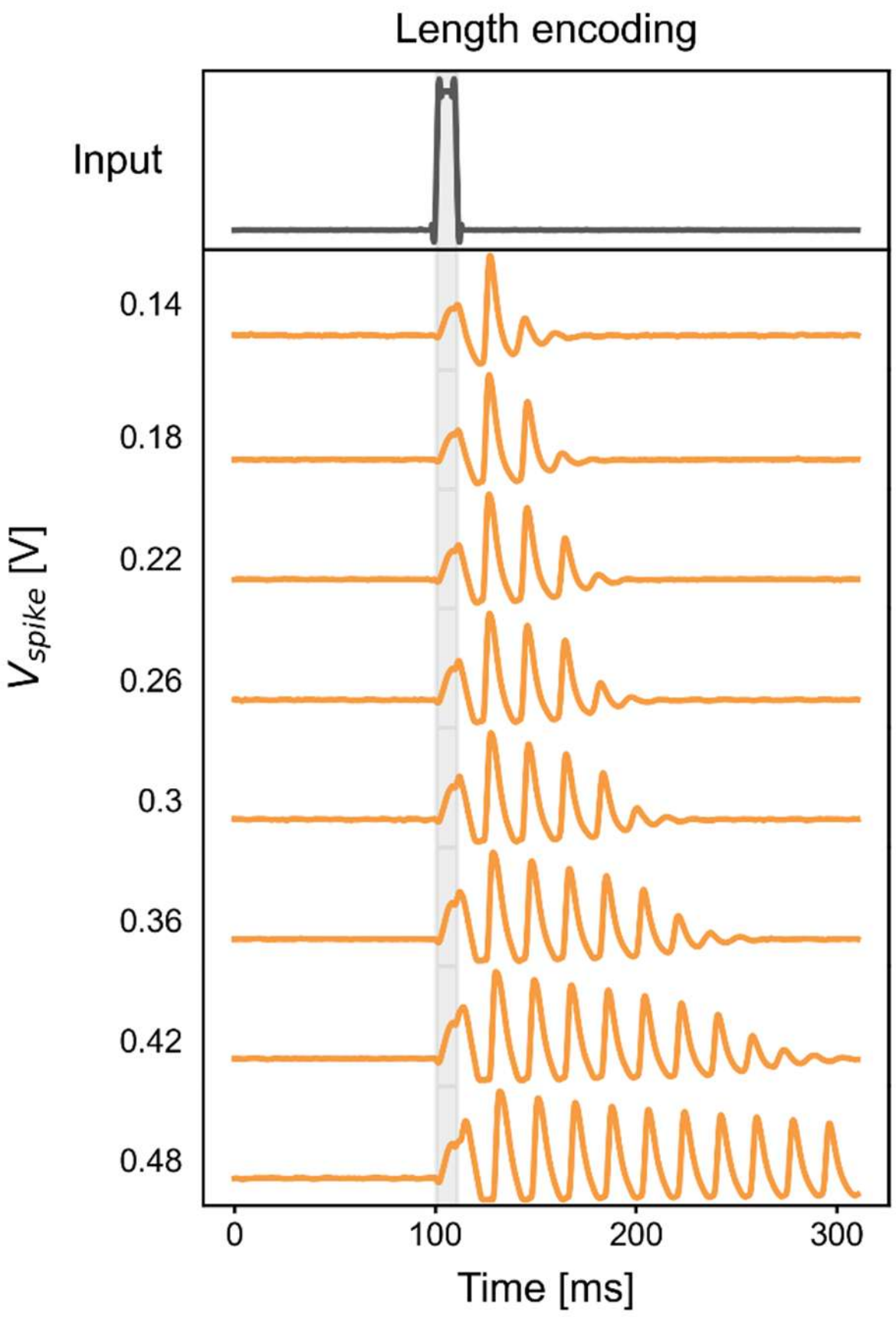

**Figure S8.** Neuron responses (yellow traces) to single input pulses (gray trace, top) for increasing $V_{spike}$ amplitudes (0.14–0.48 V, as indicated). The control voltage was kept constant at 0.80 V. Increasing $V_{spike}$ results in a progressive extension of the output burst length.

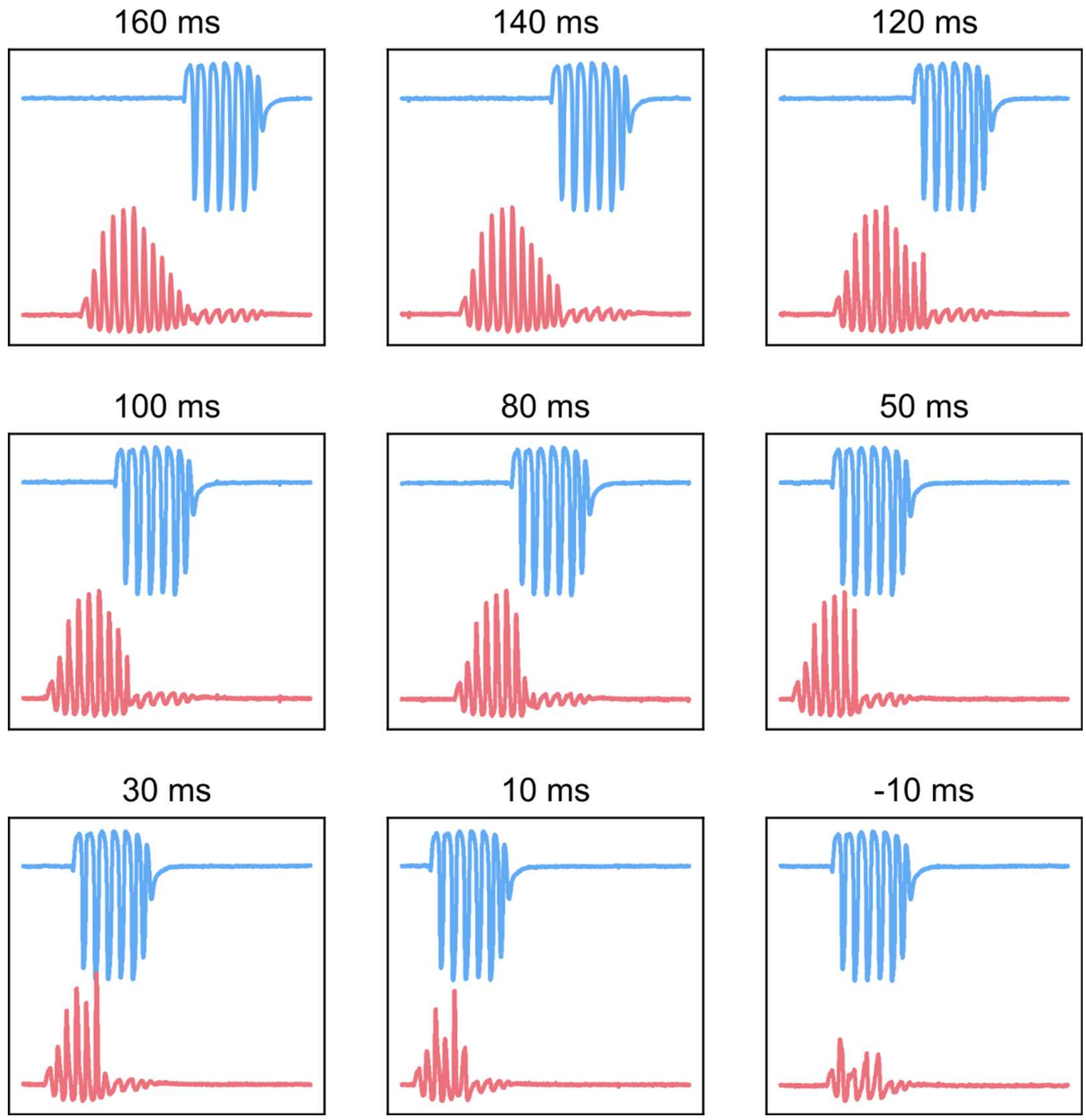

**Figure S9.** Representative temporal responses of the readout neuron for different delays $\Delta t$ between excitatory and inhibitory stimuli, as indicated above each panel (from 160 ms to −10 ms). The upper (blue) traces correspond to the inhibitory neuron output, and the lower (red) traces show the readout neuron response. As $\Delta t$ decreases, the neuron response transitions from amplification to propagation and eventually suppression, highlighting the delay-dependent modulation of spike generation.

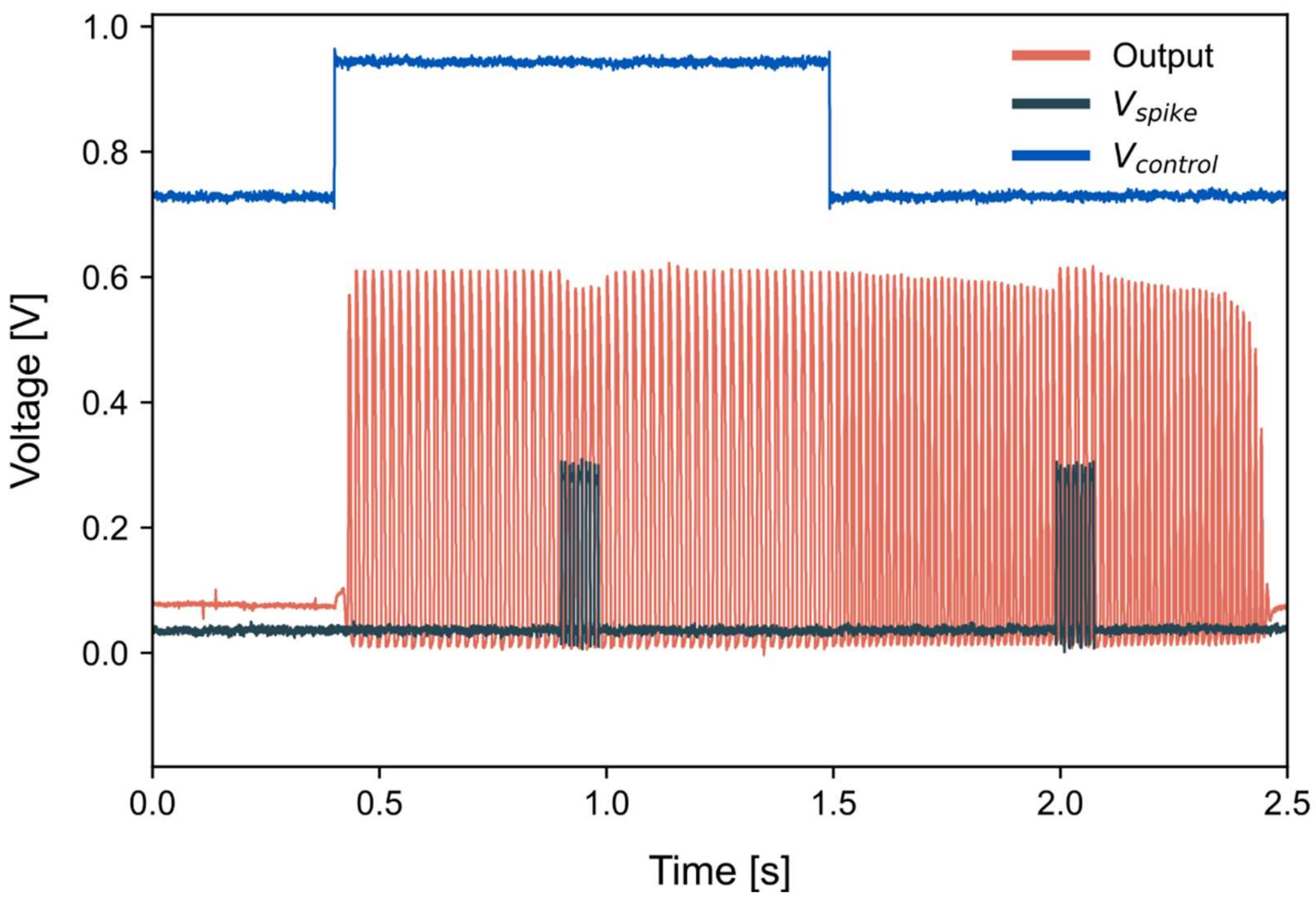

**Figure S10.** Continuous spiking activity (orange trace) is induced by an increased control voltage, $V_{control}$ (blue trace), and persists independently of the incoming input pulses (gray trace), indicating self-sustained activation under strong biasing conditions. The highly excited state persists after the 1 s stimulation offset due to the high applied voltage.

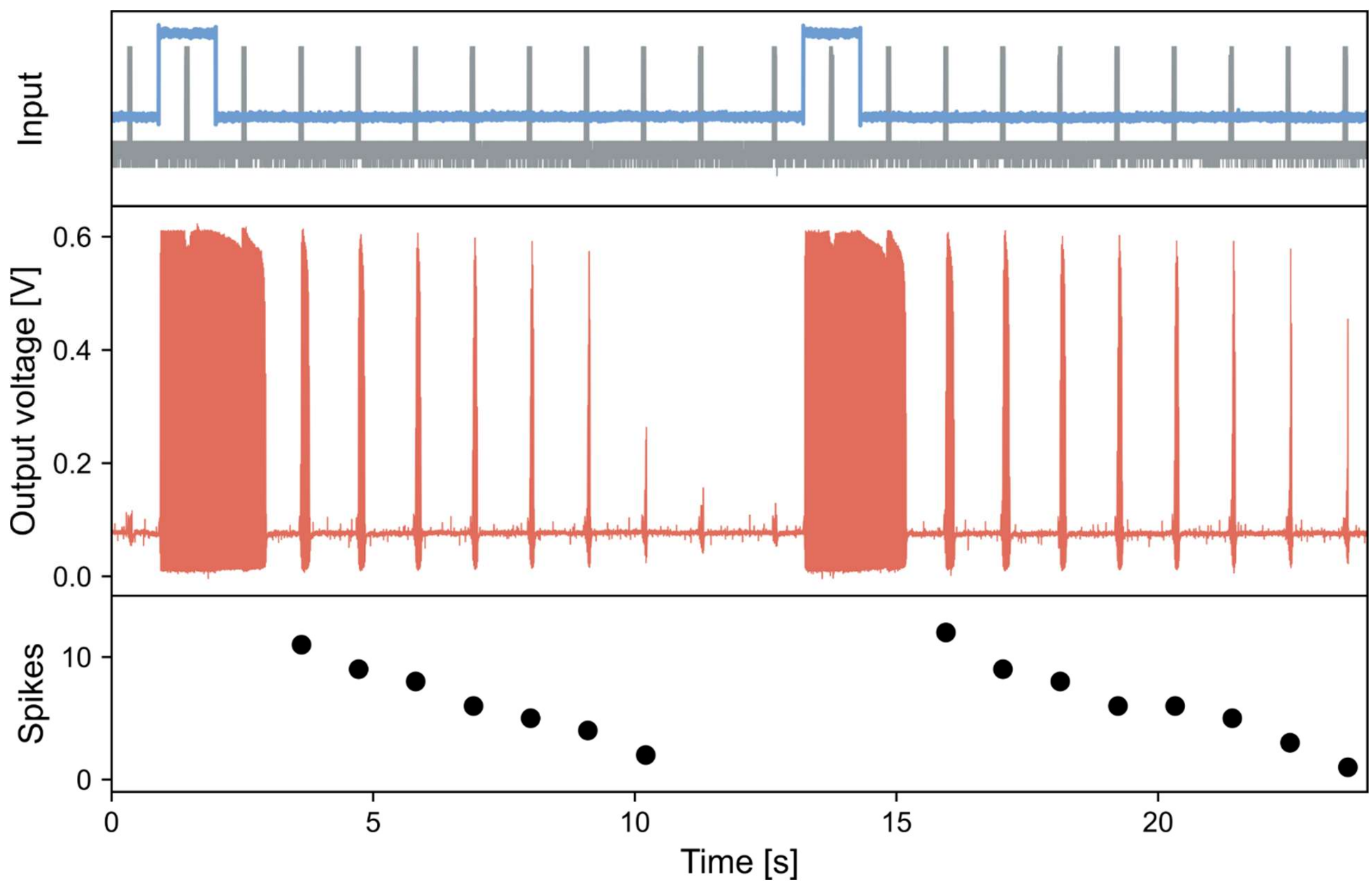

**Figure S11.** Two cycles of temporal response of the neuron (orange trace) to repetitive stimulation with sets of five 0.3 V pulses (gray trace), recorded before and after conditioning with a strong stimulus (green trace), demonstrating a reproducible short-term memory effect. Bottom panel shows number of spikes in each consecutive response.

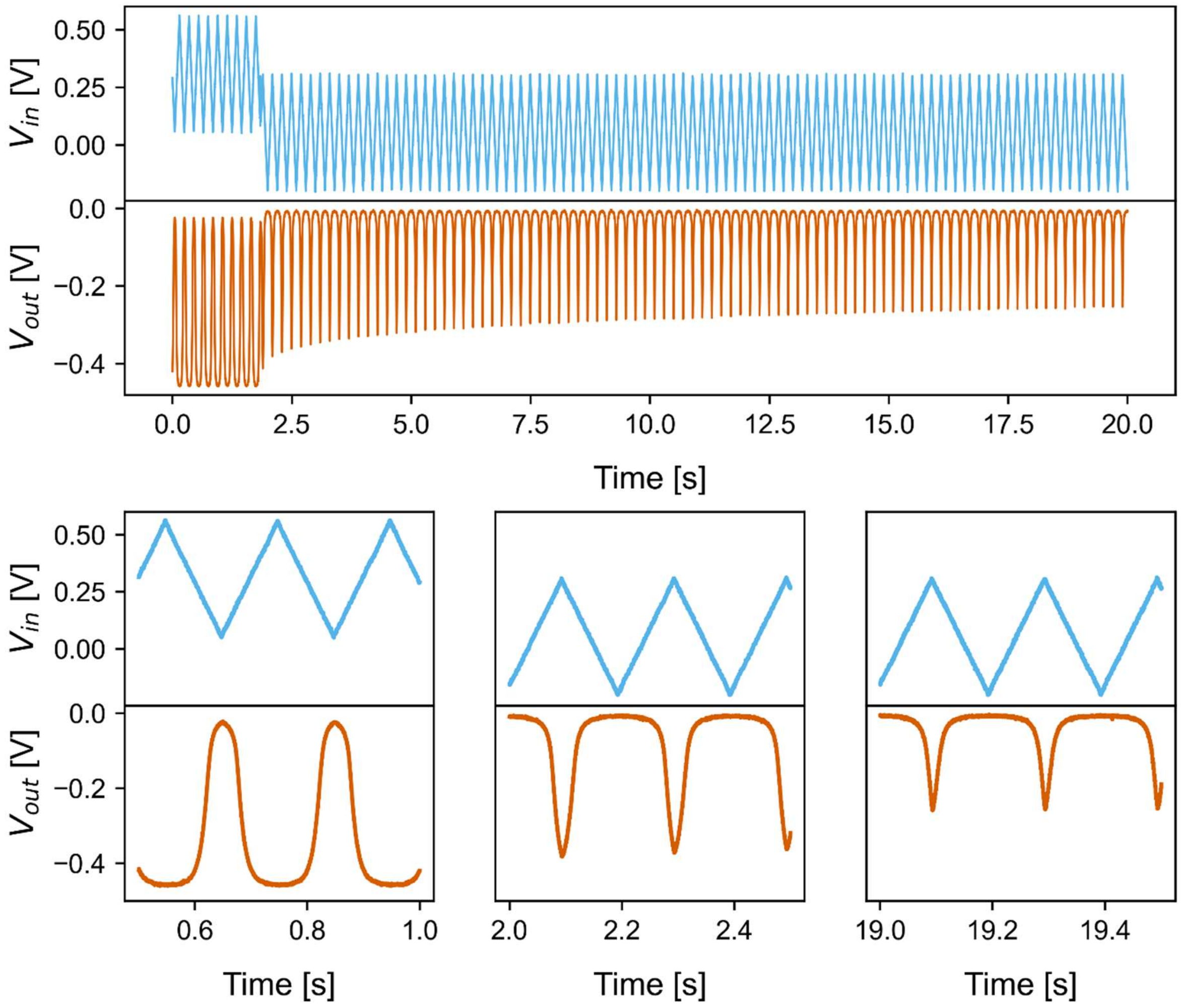

**Figure S12.** Progressive reduction in the output voltage range and peak amplitude of the OECT inverter is observed (orange trace) after lowering the input offset while maintaining the same input voltage range (blue trace). The bottom panels present magnified views of the main (top) plot.

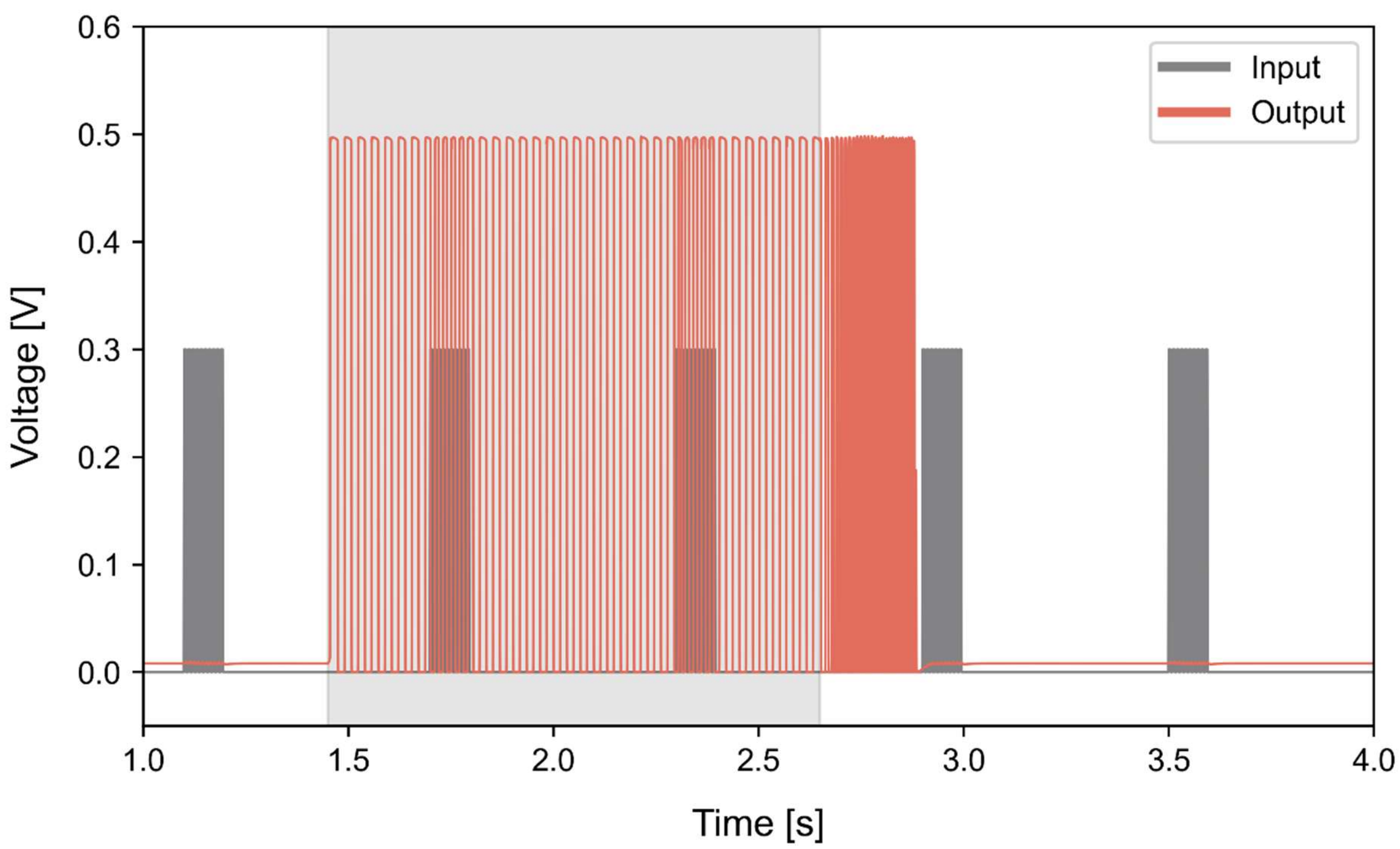

**Figure S13.** SPICE simulations of the short-term memory protocol performed with MOSFET-based circuitry show no gradual suppression of the response under repeated stimulation. In contrast to the OECT implementation, the output remains stable across identical input pulses, indicating the absence of intrinsic short-term memory effects in purely electronic devices.

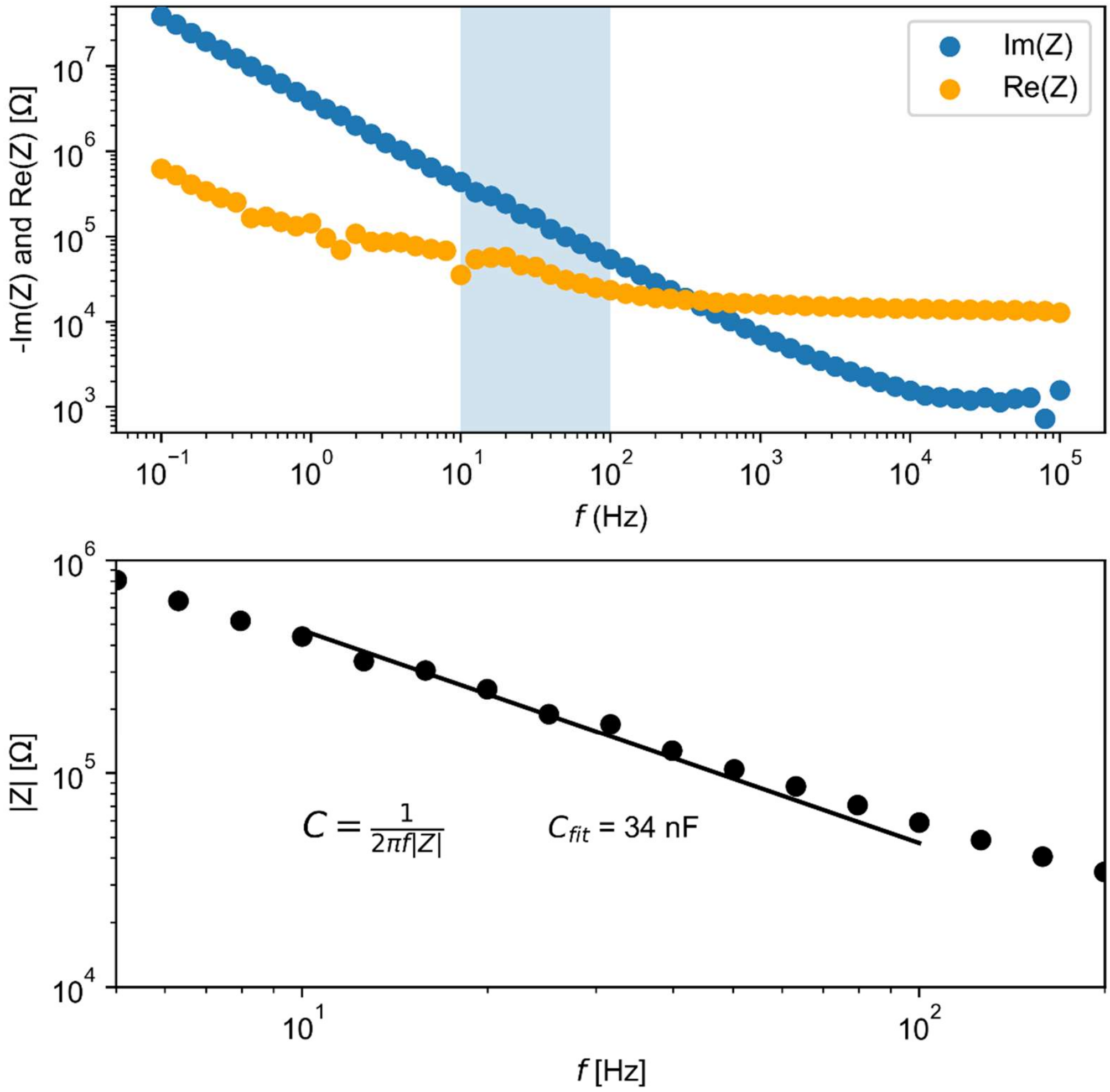

**Figure S14.** Impedance spectrum of PEDOT:PSS capacitor. The top panel shows the imaginary and real components of the impedance, with the real part remaining significantly smaller within the frequency range of interest (10–100 Hz, blue area). The bottom panel presents the magnitude of the impedance (circles) within this range and its fit (solid line) obtained using the provided equation, yielding a capacitance value of 34 nF.

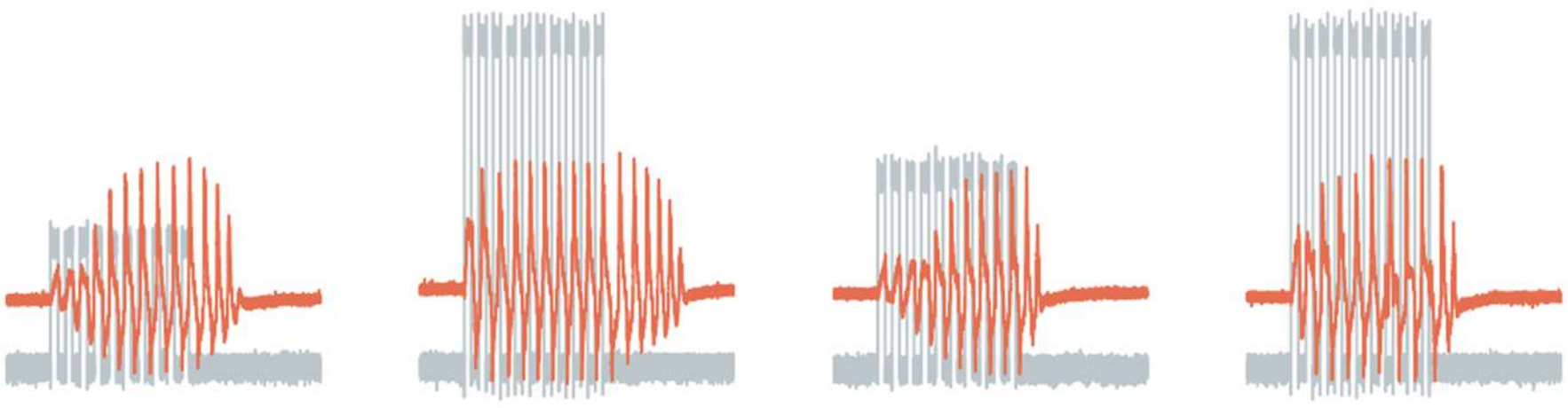

**Figure S15.** Representative responses of the fully integrated neuron to different $V_{control}$ (from left to right: 0.7 V, 0.7 V, 0.55 V, 0.55 V) and $V_{spike}$ (from left to right: 0.2 V, 0.5 V, 0.3 V, 0.5 V). In this configuration, the dendrite serves as a constant $R_{spike}$ resistor of 3 kΩ to deliver input pulses to the neuron, while a second input voltage $V_{control}$ is applied through a conventional 3 kΩ resistor.

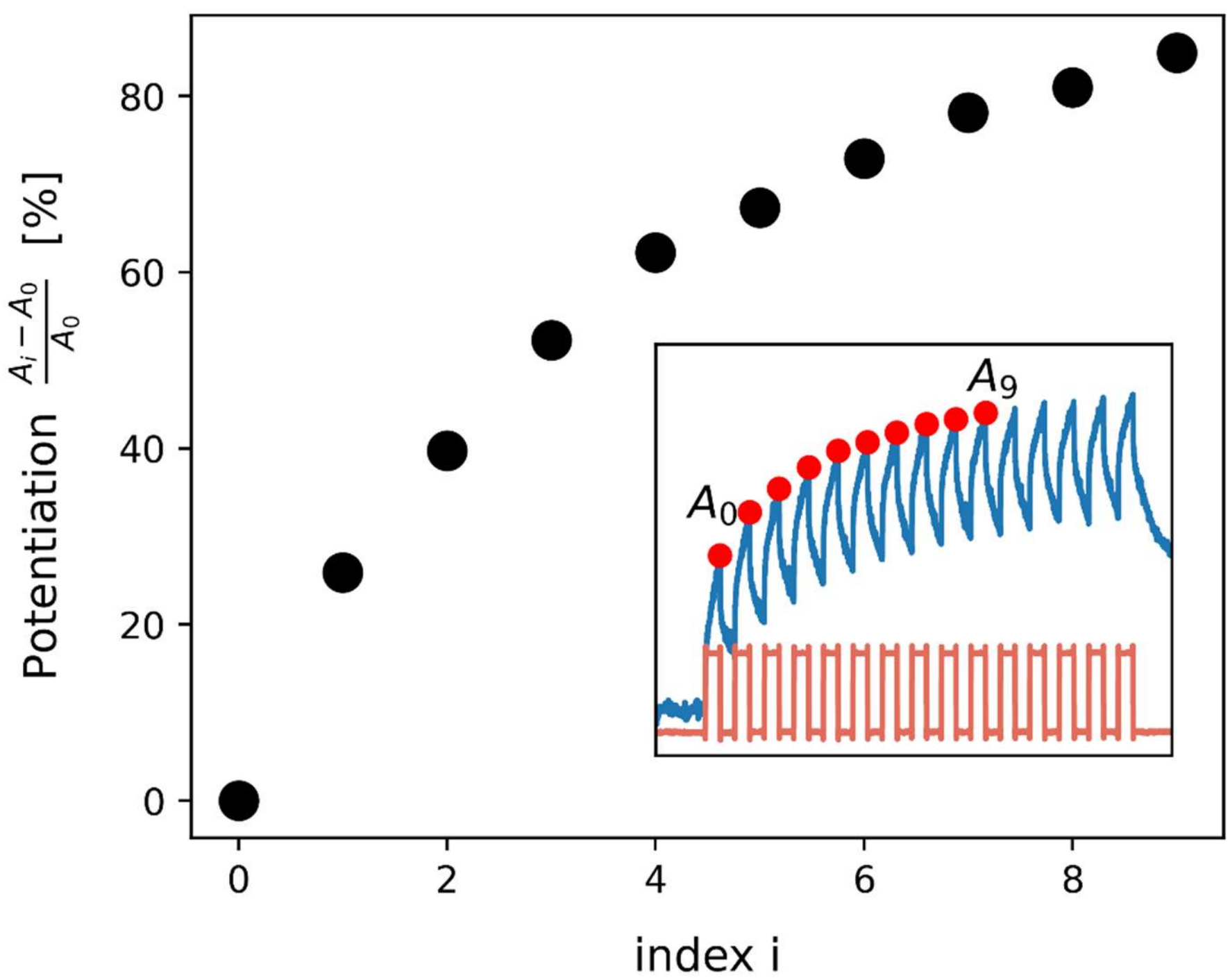

**Figure S16.** Relative potentiation of a PEDOT:$PF_6$ dendritic structure induced by consecutive pulses (0.5 V, 50 Hz). The main plot (black circles) shows the relative increase in voltage drop across the dendrite, connected in series with a 100 kΩ resistor, measured at the end of each *i*-pulse. The inset shows the input signal (red trace) and voltage drop across the dendritic structure (blue trace) with amplitudes used for calculation of relative potentiation highlighted (red dots).